# Experimental and numerical determination of mechanical properties of polygonal wood particles and their flow analysis in silos

Fernando Alonso-Marroquín[1*], Álvaro Ramírez-Gómez[2], Carlos González-Montellano[2], Nigel Balaam[1], Dorian A.H. Hanaor[1], E.A. Flores-Johnson[1], Yixiang Gan[1], Shumiao Chen[1], and Luming Shen[1]

1   School of Civil Engineering, The University of Sydney, Sydney, NSW, Australia
2   BIPREE Research Group, Universidad Politécnica de Madrid, Madrid, Spain:
    *Corresponding author: fernando.alonso@sydney.edu.au

**Abstract**  Responding to a lack in the literature, mechanical properties of polygonal wood particles are determined for use in a discrete element model (DEM) for flow analysis in silos, and some methods are proposed for determining such parameters. The parameters arrived at here have also formed part of the input to the SPOLY software, developed in-house to compute the DEM model with spheropolyhedron elements. The model is validated using a 2D physical model, where "prismatic" particles with polygonal cross sections are placed inside a silo with variable aperture and hopper angle. Validation includes comparison of flow-rates computed by SPOLY, displacement profiles, and clogging thresholds with experimental results. The good agreement that emerges will encourage future use of miniature triaxial tests, grain-surface profilometry, inclined slope tests, and numerical analysis of the intragranular stresses – toward a direct construction of the contact-deformation relations required in realistic DEM modelling of particle flow with angular-shaped particles.

## 1  Introduction

DEM models for flow studies in silos have normally assumed spherical particles [6, 23, 25, 36]. Aiming at a better approximation to what industries actually confront, some also consider multi-spherical particles [9, 16, 17, 19], most of which present non-spherical geometries (ellipsoidal, ovoid, pointed shape). Increasing the number of sphere components improves the approximation to real particles; but despite such refinements, there will always be a waviness effect that needs to be checked. While particles with angular geometries are most distant from the assumed spherical ideal, such particles are frequently encountered in mining, and more recently in biomass industries. Further studies are needed to achieve reliable storage and handling of solid biofuels from woodchips, for example [7, 24, 29, 30].

Several DEM models have been proposed to simulate angular particles in 2D and 3D. DEM using Voronoi polygons [33] was successfully used in applications with granular matter [2]. Peña et al. [21] and Hidalgo et al. [20] used a 2D approach to study the packing properties of rods (with varying elongation) settling under gravity. But these methods have serious limitations: they were not easily extended to 3D, and they only allow one contact per pair of particles. For greater accuracy, multiple contacts must be simulated. In particular, these can create contact moments that are responsible for stability of arches and buckling of force chains [34]. Alonso-Marroquin [4] proposed to combine the idea of the Minkowski sum approach [32] with multiple-contact laws to model the realistic interactions of complex shapes. This model was later used to simulate angular particles using both Voronoi-Minkowski diagrams [13] and Voronoi spheropolyhedra [14]. Galindo-Torres et al. [12] introduced 3D molecular dynamics (MD) techniques using spheropolytopes, and defined a multi-contact law for two bodies that allowed simulations with a wide range of particle shapes. Using spheropolygons, Kanzaki et al.

[22] undertook a systematic theoretical and experimental study of the structural and mechanical properties of the packing of faceted particles, after their partial discharge from a silo. Hidalgo et al. [21] investigated the formation of ordered structures in cohesive particles using spherocylinders. Acevedo et al. [1] used spheropolygons to explore the effect of the pouring mechanism on the structural properties of deposits of square particles in a rectangular silo. The problem of finding parameters for such a DEM model – along with its validation – remained unsolved.

This paper reports the validation of a two-dimensional model using spheropolyhedra, through experiments with angular wood particles in a two-dimensional silo. The validation was based on a direct determination of the parameters of the DEM model. Miniature triaxial tests were used to calculate contact stiffness; pendulum tests and drop tests were carried out to determine restitution coefficients; and inclined slope tests were conducted to determine coefficients of friction.

The paper is organized as follows: Section 2 presents the DEM model; Section 3 the determination of DEM parameters; Section 4 the validation and the analysis of the sensitivity of the numerical results with the DEM parameters; and Section 5 the conclusions.

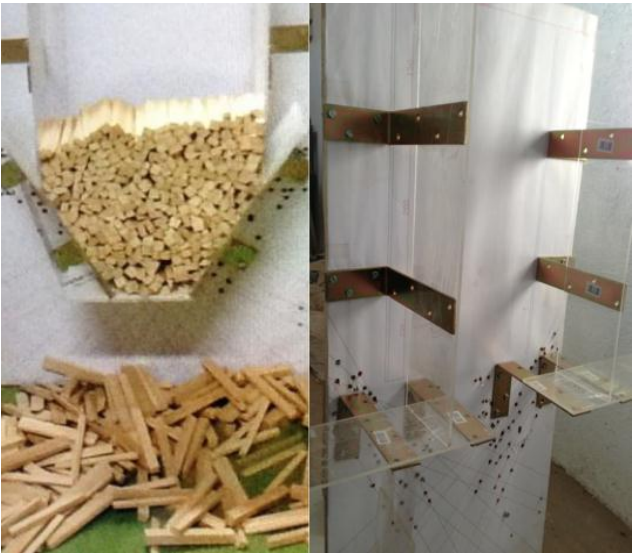

**Figure 1.1 Two-dimensional model silo. Front view (left), silo filled with wood particles and a hopper configuration; and side view (right), silo empty of particles and a flat-bottom silo configuration.**

## 2 The SPOLY model

The DEM used in this paper was based on the team's in-house software: SPOLY, an object-oriented C++ program that tracks particles and interactions using high-order explicit solvers of the equations of motion. Particle shapes were modelled using the concept of the *spheropolyhedron* – the Minkowski sum of a polyhedron and a sphere [5]. Recently a graphical interface called PREPS has been implemented to allow rapid construction of models without needing to write code. The structure of the SPOLY code has been presented before [5]; here we present only the contact-interaction model used in the granular flow simulations of this paper.

Within SPOLY, interparticle interactions are calculated using vertex-face and edge-edge interactions. A special case was when the polyhedra had uniform cross section and their kinematics were restricted to two-dimensional displacement and rotation along their principal axis only. In this case the interparticle interactions were calculated using vertex-edge contacts between the polygonal cross sections. Each contact force was calculated as

$$\vec{F}^c = \vec{F}^e_n + \vec{F}^e_t + \vec{F}^v_n + \vec{F}^v_t, \qquad (2.1)$$

where the elastic forces were given by

$$\vec{F}^e_n = -k_n \Delta x_n \vec{n} \quad \vec{F}^e_t = -k_t \Delta x_t \vec{t} \qquad (2.2)$$

and $\vec{n}$ and $\vec{t}$ are the normal and tangential unit vectors. The scalar $\Delta x_n$ is the overlapping length: the vertex-to-edge distance between two particles. The scalar $\Delta x_t$ accounts for the tangential elastic displacement given by the frictional force, and it satisfies the sliding condition by $\left| F^e_t \right| \leq \mu F^e_n$, where $\mu$ is the coefficient of friction. Here, $k_n$ and $k_t$ are the normal and tangential coefficients of stiffness. The last two terms on the right hand side of Eq. (2.1) account for energy loss after collision. They are calculated as

$$\vec{F}_n^v = -m\gamma_n v_n \vec{n} \quad \vec{F}_t^v = -m\gamma_t v_t \vec{t}, \qquad (2.3)$$

where the effective mass is $m = m_1 m_2/(m_1+m_2)$ and the mass of the particle is $m_i = \rho A_i$ (i=1,2). The density is $\rho$, and $A_i$ represents the area of the particle. The normal and tangential coefficients of damping are given by $\gamma_n$ and $\gamma_t$ respectively; and $v_n$ and $v_t$ denote the normal and tangential components of the contact velocity.

# 3 Determination of DEM parameters

The parameters of the DEM model are:

| | |
|---|---|
| µ | coefficient of friction |
| $k_n$ and $k_t$ | elastic parameters |
| $\gamma_n$ and $\gamma_t$ | damping parameters |
| $\rho$ | surface density |

The experimental procedures used to obtain the values of these parameters are discussed in the following sections.

## 3.1 Coefficients of friction (µ)

### 3.1.1 Particle-wall coefficient of friction

The determination of particle-wall friction was based on a sliding test similar to the one described in [8]; see Figure 3.1. A block was placed on top of a methacrylate wall sample. The wall sample was then inclined until the particle on the top slid, and at that moment the angle of inclination α was recorded. Forces acting on the block, both normal and tangential, were $N = mg\cos\alpha$, $F = mg\sin\alpha$, and $mv\gamma_t$. Assuming that the friction force is given by $F = \mu N$, we can obtain $\mu = \tan\alpha$.

To obtain a representative value, tests were repeated twelve times with three particle-wall pairings. The coefficient of friction between the dry sample of wood particles and the wall ranged from 0.31 to 0.58, and a coefficient of 0.44 was derived as the overall mean. Mean values and standard deviations were determined from twelve repetitions for each sample: 0.46 (10%), 0.47 (13%), and 0.40 (23%).

### 3.1.2 Particle-particle coefficient of friction

To determine the friction between two particles, one particle was placed on top of another, whose inclination was increased until the top particle slid. The measured heights $H_1$ and $H_2$, together with the length L, allowed the angle of inclination α to be determined (see Figure 3.1).

As in the previous experiment, tests were repeated up to ten times with three pairs of particles to obtain a representative value.

The coefficient of friction between two dry samples of wood particles ranged from 0.33 to 0.43, depending on the surface of the particles. A friction coefficient of 0.38 was selected as the overall mean value. Mean values and standard deviations of the coefficient of friction for the three samples − with ten repetitions for each − were 0.38 (7%), 0.40 (7%), and 0.33 (4%).

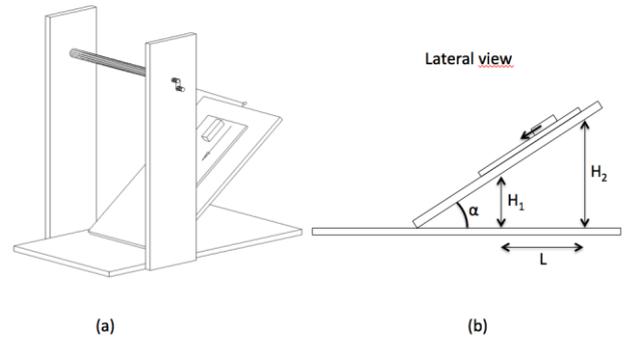

**Figure 3.1** Apparatus used to determine the particle-to-wall and particle-to-particle coefficients of friction; general view (a) and detailed view (b).

## 3.2 Constant stiffness

We proposed to obtain the normal and tangential coefficients of stiffness, $k_n$ and $k_t$ respectively, using a miniature triaxial test of two particles, as shown in Figure 3.2. The contact between the two particles has an angle α with respect to the horizontal such that $\tan(\alpha) < \mu$. The two particles were quasistatically loaded in the vertical direction, so that damping forces were absent. Deformations were tracked using marks on the particles, and the axial load was recorded.

The constant coefficients of stiffness are given by the ratio between the load applied at the contact and the resulting deformation of the sample:

$$k_n = \frac{F_n^e}{\Delta x_n} \qquad k_t = \frac{F_t^e}{\Delta x_t} \qquad (3.1)$$

Initial evaluation of stiffness coefficients was carried out using equilibrium equations of the contact forces and the measured axial force applied to the particles, as shown in Figure 3.2 (bottom). A non-linear force displacement was observed. However, this curve could not be used to obtain the constant stiffness, owing to particle rotation for any α>0. This suggested the existence of a contact moment between the particles that was not accessible in our uniaxial experiment. To overcome this difficulty, we first calculated $k_n$ by taking α=0. Then $k_t$ was obtained using numerical calculation of the intra-granular stress taking α>0, and using finite element modelling as described below.

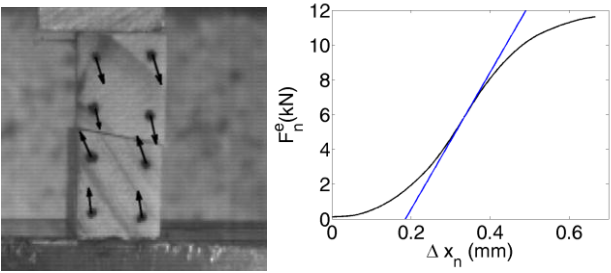

**Figure 3.2 (left) Miniature triaxial test used to measure load-displacement relation; (right) between two wood particles.**

### 3.2.1 Normal stiffness ($k_n$)

The normal stiffness was obtained by uniaxially loading two particles perpendicular to the contact surfaces – in both longitudinal and transverse directions relative to the wood fibres, using cubic specimens of 10mm vertex length. Because the wood's deformation was non-linear, stiffness was determined from loading-unloading cycles (see [2]). Stress-strain curves with unloading-stiffness interpolations are shown in Figure 3.3; and the topography of particle surfaces is illustrated in Figure 3.4.

For consistency, measurements were repeated five times each in transverse and longitudinal directions, yielding a transverse compressive modulus of 242 MPa (standard deviation 15%) and a longitudinal compressive modulus of 650 MPa (standard deviation 33.5%). Values for the uniaxial compressive modulus were found to be significantly lower than values for the tensile modulus of wood, typically reported to be around 10 GPa. Since the wood fibres were perpendicular to the cross-section area of the particles and only transversal loads were applied, we employed the mean transverse modulus to determine stiffness values. Using $k_n$=EA/L, with average particle dimensions A=10mm×120mm and L=20mm, the representative stiffness value we obtained was kn=(1.00±0.18)×10⁴ N/mm and the peak value was kn=(3.9±0.18)×10⁴ N/mm.

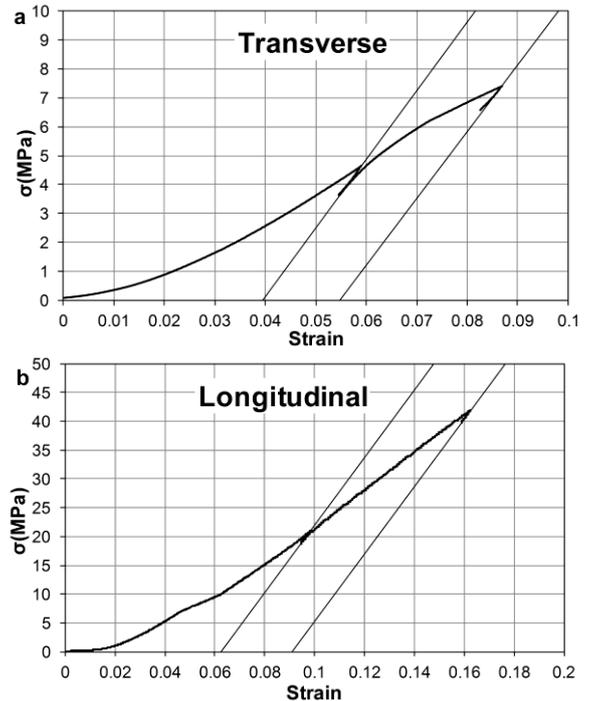

**Figure 3.3 Stress versus strain in loading-unloading cycles in (top) transverse direction and (bottom) longitudinal direction, relative to wood fibres.**

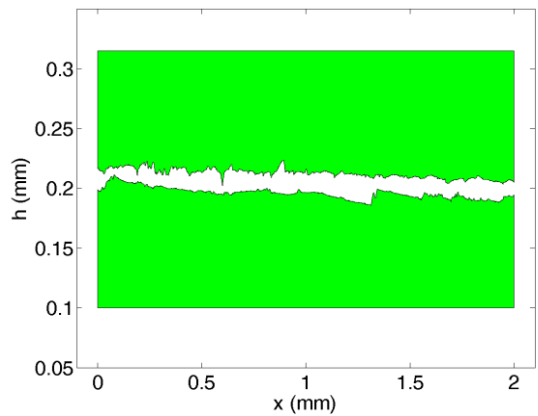

**Figure 3.4 Topography of particle surfaces: top line shows transverse direction (perpendicular to fibres); and bottom line, longitudinal (parallel to fibres).**

Non-linearity in the stress-strain behaviour of particles in compression, along with large standard deviations on the measure of $k_n$ and $\mu$, stems in part from the roughness of the contacting surfaces. Surface profiles, measured by stylus profilometry, exhibited amplitudes of 100μm transversally and ~200μm longitudinally (Figure 3.4). This is partly due to striations from sawing the material. Surface roughness makes wood-wood contact softer, effectively decreasing the Young modulus for the bulk material. The flattening of contact asperities was evident from gradual increases in stiffness during early stages of loading. Owing to the quasi-randomness of surface roughness, the alignment of surface asperities at particle contacts differs for each initial configuration. This explains the large standard deviation of $k_n$ and $\mu$, and may manifest in the variability of macroscopic discharge flow, with greater significance for smaller systems with fewer interparticle contacts.

We do expect roughness to influence the contact stiffness parameters and the coefficient of friction. Yet we were able to measure $k_n$ and $\mu$ directly from experiments, without dealing with the roughness. Measuring $k_t$ is possible if one could design the miniature triaxial test in such a way that contact moments are absent. But it was most feasible to measure experimentally the main features of roughness, and to use a finite element analysis method for $k_t$, as detailed in the next section.

### 3.2.2 Tangential stiffness $k_t$

The tangential stiffness is calculated from the intergranular fields. The elastic force around a contact can be calculated in terms of the intragranular stresses as

$$F_n^e = \sigma_n A \qquad F_t^e = \sigma_t A \ , \qquad (3.2)$$

where $\sigma_n$ and $\sigma_t$ are the normal and shear stresses at the contact interface, and A is the area of the contact. Substituting Eqs. (3.2) into Eqs (3.1) we obtain

$$\frac{k_t}{k_n} = \frac{\sigma_t}{\sigma_n} \frac{\Delta x_n}{\Delta x_t} \ . \qquad (3.3)$$

The intergranular fields were calculated for the blocks in Figure 3.2 using the finite element program ABAQUS/Explicit. The blocks were modelled as an elastic material with Poisson ratio of $\nu=0.2$ and Young modulus of E=242 MPa. To reproduce the displacement shown by the markers in Figure 3.2, we modelled the main features of the roughness of the interface and introduced atomic friction of $\mu_a = 0.2$ (see [26]). The rough surface used in the model is shown in Figure 3.5. Six markers, shown in Figure 3.6, were used to calculate the stresses and deformation of the blocks.

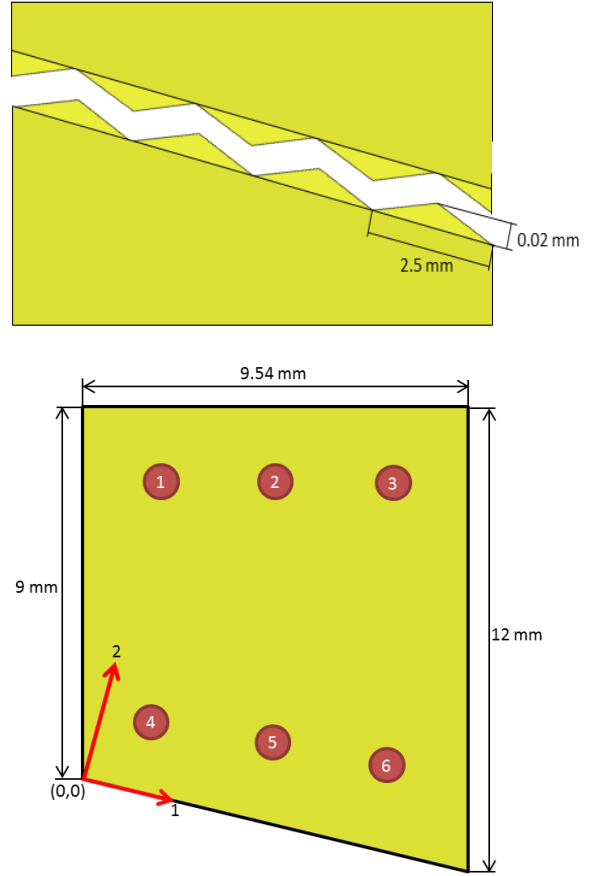

**Figure 3.5 (top)** Model for rough surface used in the finite element calculation; (bottom) markers used to calculate stress and displacements as listed in Table 3.1.

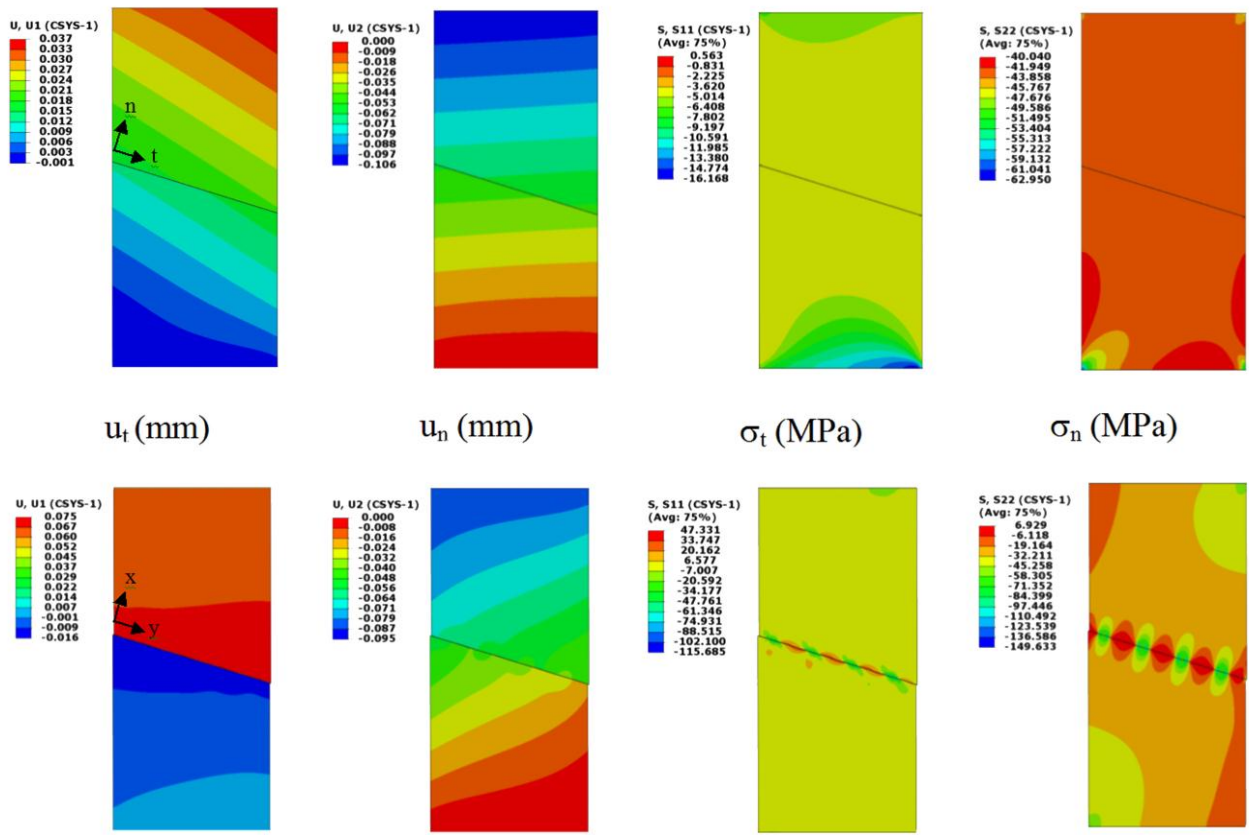

**Figure 3.6 Top images are for flat surfaces with friction coefficient of 0.37; bottom images are for rough surfaces with friction coefficient of 0.2. Pairs from left to right: tangential displacement, normal displacement, tangential stress, and normal stress.**

Stress and displacement values are shown in Figure 3.6 for two different models. In the first, the contact was modelled as a "flat surface" with coefficient of friction µ=0.37. In the second model, it was "rough surface" with atomic friction coefficient 0.2. The deformation was quite uniform in the flat surface model, but discontinuous at the contact region in the rough surface model, agreeing with the experimental displacement shown in Figure 3.2. The stress in the blocks was quite uniform in the flat surface model, while in the rough surface it was concentrated at the interface. In the rough surface model the normal stress fluctuates strongly, alternating between tensile and compressive values.

The six markers shown in Figure 3.5 (top) were used to calculate the stresses and displacements shown in Table 3.1. The normal stress in the rough surface model is significantly lower than for the flat surface, showing that the rough surface decreases normal stiffness at the contact, typically by 50%. Tangential stiffness was calculated for both models using Eq. (3.3). The resulting values are $k_t = (2.8 \pm 0.09)k_n$ for the flat surface and $k_t = (0.065 \pm 0.04)k_n$ for the rough surface. The ratio of tangential to normal stiffness is significantly lower for flat surfaces than for rough surfaces. Stresses in the proximity of rough surfaces fluctuate significantly, depending on the surface profile.

### 3.3 Normal coefficient of restitution

The normal coefficient of damping $\gamma_n$ was obtained using a pendulum collision test (see [35]). The coefficient of restitution represents the degree of conservation of kinetic energy after collisions between particles (particle-to-particle coefficient of restitution $\varepsilon_p$) or between particles and the silo wall (particle-to-wall coefficient of restitution $\varepsilon_w$). Its value is derived from the kinetic energy of the particle before and after the collision. When particles are not subject to rotation, the coefficient of restitution is obtained from Eq. (3.4).

| Flat surface | | | | | | | |
|---|---|---|---|---|---|---|---|
| | Point | 1 | 2 | 3 | 4 | 5 | 6 |
| Displacement $\Delta x_t$ (mm) | | 0.03 | 0.03 | 0.03 | 0.02 | 0.02 | 0.02 |
| Displacement $\Delta x_n$ (mm) | | -0.09 | -0.03 | -0.09 | -0.07 | -0.07 | -0.06 |
| Stress $\sigma_t$ (MPa) | | -4.53 | -4.64 | -4.30 | -4.18 | -4.14 | -4.18 |
| Stress $\sigma_n$ (MPa) | | -42.62 | -42.71 | -42.75 | -42.57 | -42.65 | -42.55 |
| Coordinate 1 (mm) | | 0.18 | 2.86 | 5.64 | 1.64 | 4.58 | 7.57 |
| Coordinate 2 (mm) | | 6.46 | 7.30 | 8.17 | 1.12 | 1.79 | 2.52 |
| $k_t/k_n$ | | 0.37 | 0.11 | 0.29 | 0.36 | 0.31 | 0.27 |
| Rough surface | | | | | | | |
| | Point | 1 | 2 | 3 | 4 | 5 | 6 |
| Displacement $\Delta x_t$ (mm) | | 0.06 | 0.06 | 0.06 | 0.07 | 0.07 | 0.07 |
| Displacement $\Delta x_n$ (mm) | | -0.08 | -0.08 | -0.07 | -0.06 | -0.06 | -0.05 |
| Stress $\sigma_t$ (MPa) | | -0.60 | -1.36 | -2.30 | -3.28 | -2.04 | -0.35 |
| Stress $\sigma_n$ (MPa) | | -21.65 | -28.85 | -35.11 | -20.86 | -25.80 | -28.60 |
| Coordinate 1 (mm) | | 0.18 | 2.86 | 5.64 | 1.64 | 4.58 | 7.57 |
| Coordinate 2 (mm) | | 6.46 | 7.30 | 8.17 | 1.12 | 1.79 | 2.52 |
| $k_t/k_n$ | | 0.04 | 0.06 | 0.08 | 0.14 | 0.06 | 0.01 |

**Table 3.1 Intergranular values at the six markers shown in Figure 3.5 using the model in Figure 3.6**

$$\varepsilon = -\frac{v_1 - v_2}{u_1 - u_2} \quad . \quad (3.4)$$

The subindices 1 and 2 refer to the elements involved, and $u$ and $v$ to the velocities just before and after the collision.

### 3.3.1 Particle-to-wall coefficient of restitution

This coefficient is obtained from a drop test (see [8, 10, 11, 18, 35]) as shown in Figure 3.7. Our test involved the controlled fall of a particle against a flat methacrylate surface, recorded at 100 frames/second. We used an Infaimon high-speed Genie H1400 monochrome camera. Each of three particles was dropped ten times from three different heights (90 assays). The height reached after bouncing was obtained by image analysis.

The velocities $v_2$ and $u_2$ in (Eq. 3.4), corresponding to the flat surface, are considered to be zero. Assuming that energy is conserved before and after impact, the resulting particle-to-wall restitution coefficient is

$$\varepsilon_w = -\frac{v_1}{u_1} = \sqrt{\frac{H_1}{H_0}} \quad . \quad (3.5)$$

The mean values and standard deviation obtained from the experiments are summarized in Table 3.2

| | $H_0$ (mm) | 18.92 | 44.55 | 45.48 |
|---|---|---|---|---|
| **Particle A** | mean | 0.85 | 0.60 | 0.59 |
| | std | 2% | 4% | 3% |
| **Particle B** | mean | 0.39 | 0.42 | 0.48 |
| | std | 6% | 12% | 7% |
| **Particle C** | mean | 0.47 | 0.56 | 0.42 |
| | std | 10% | 8% | 13% |
| **Average** | mean | 0.56 | 0.53 | 0.49 |
| | std | 35% | 17% | 16% |
| **Mean** | | | | 0.53 |

**Table 3.2** Mean values of the particle-wall restitution coefficient. Experiment was repeated 10 times using three particles (A, B, and C) and three different initial heights $H_0$. The final height $H_1$ was used to calculate the restitution coefficient using Eq. (3.5).

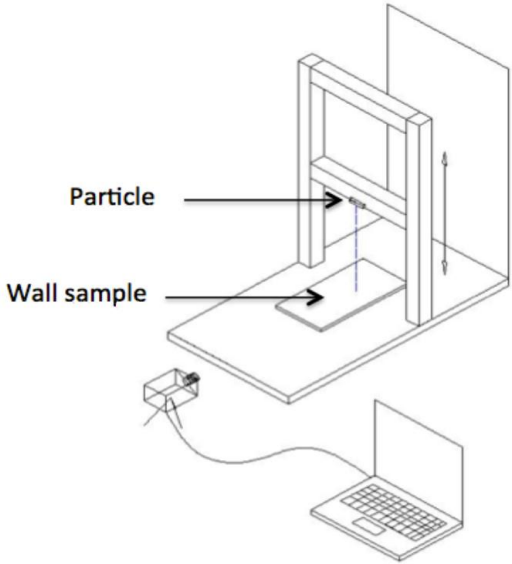

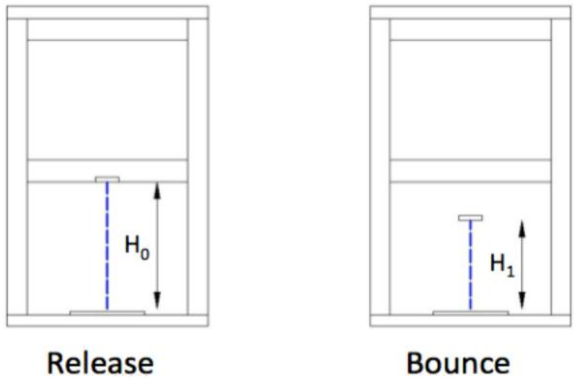

**Figure 3.7** Apparatus used to determine the particle-wall coefficient of restitution.

### 3.3.2 Particle-to-particle coefficient of restitution

The coefficient of restitution between particles was determined using two identical blocks (compare [34]) suspended on a double pendulum (Figure 3.8). They were perfectly aligned – glued to, and kept horizontal by, nylon strings of equal length. Three release heights ($H_0$) were used. When particle 1 was released and impacted against particle 2, the movements and heights of both ($H_1$ and $H_2$) were recorded using the high-speed camera.

Considering that the velocity $u_2$ is zero in Eq. (3.4) and that the energy of the particles before and after the impact is conserved, the non-zero velocities in Eq. (3.4) can be expressed as a function of the heights $H_1$ and $H_2$. Therefore, the value of $\varepsilon_p$ is given by

$$\varepsilon_p = -\frac{v_1 - v_2}{u_1} = \frac{\sqrt{H_2} - \sqrt{H_1}}{\sqrt{H_0}} \quad . \quad (3.6)$$

Each test was repeated 15 times. The mean value and standard deviation are included in Table 3.3.

| $H_0$ (mm) | 69.39 | 91.27 | 111.14 |
|---|---|---|---|
| mean | 0.47 | 0.47 | 0.41 |
| std | 18% | 6% | 22% |
| **Mean** | | | **0.45** |

**Table 3.3** Mean values of the particle-particle restitution coefficient. The experiment was repeated 12 times using the pendulum test with three different initial heights $H_0$. The final heights $H_2$ and $H_1$ (which was almost zero) were used with Eq. (3.6) to calculate the restitution coefficient.

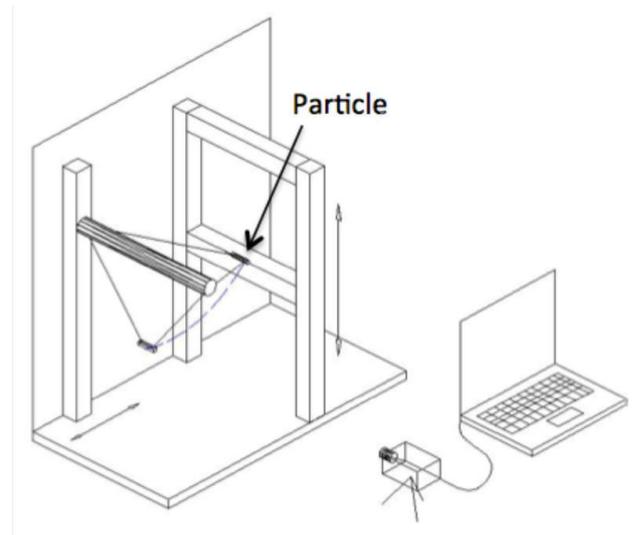

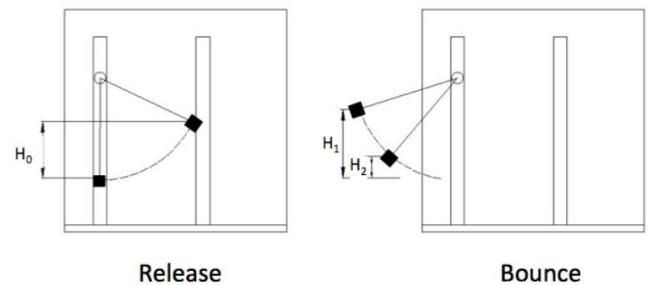

**Figure 3.8** Apparatus used to determine the particle-to-particle coefficient of restitution.

### 3.3.3 Normal coefficient of damping $\gamma_n$

The coefficients of restitution can be related to the normal coefficient of damping $\gamma_n$ using an analytical derivation of the collisions between two blocks (compare [28]). The penetration depth $\delta = \Delta x_n$ of two blocks of mass $m_1$ and $m_2$ satisfy the differential equation

$$\frac{d^2\delta}{dt^2} + \gamma_n \frac{d\delta}{dt} + \omega_0^2 \delta = 0 \quad . \qquad (3.7)$$

This equation is solved with initial condition $\delta=0$ and $d\delta/dt=v_0$ at $t=0$, where $v_0$ is the relative velocity before impact (see Appendix). If $v_f$ is the relative velocity after impact, the analytical solution for the restitution coefficient is

$$\varepsilon = \frac{-v_f}{v_0} = \frac{v_0 \exp\left(-\frac{\gamma_n}{2}\frac{\pi}{\omega}\right)}{v_0} = \exp\left(-\frac{\gamma_n}{2}\frac{\pi}{\omega}\right)$$

$$\omega = \sqrt{\omega_0^2 - \left(\frac{\gamma_n}{2}\right)^2} \quad \omega_0 = \sqrt{\frac{k_n}{m_{12}}} \quad . \qquad (3.8)$$

Here $m_{12}$ is the effective mass of the two particles. We use this formula to calculate $\gamma_n$ for particle-particle interaction. In this case the effective mass is $m = m_1/2$, and the formula above results in

$$\varepsilon_p = \exp\left(-\frac{\gamma_n}{2}\frac{\pi}{\sqrt{\frac{2k_n}{m_1} - \left(\frac{\gamma_n}{2}\right)^2}}\right) \quad . \qquad (3.9)$$

For the case of the particle-wall interaction the mass of the wall is much greater than the mass of the particle, so that $m = m_1$. The formula becomes

$$\varepsilon_w = \exp\left(-\frac{\gamma_n}{2}\frac{\pi}{\sqrt{\frac{k_n}{m_1} - \left(\frac{\gamma_n}{2}\right)^2}}\right) \quad . \qquad (3.10)$$

A particle-particle coefficient of restitution of 0.53 has been used, and this gives a normal coefficient of damping $\gamma_n$ of 6085 *1/s*. The particle-wall coefficient of restitution is 0.45, yielding coefficient of damping of $\gamma_n$ =4303 1/s.

### 3.4 Tangential coefficient of damping $\gamma_t$

The tangential coefficient of damping is calculated by sliding a block of mass m over another block of mass M>>m using the apparatus shown in Figure 3.1.
The inclination of the slope was $\theta > \tan^{-1}\mu$. The equation of motion of the sliding block is

$$\frac{dv}{dt} = -\gamma_t v + g\sin\theta - \mu g\cos\theta \quad . \qquad (3.10a)$$

The solution of this equation is

$$v = v_0(1-e^{-\gamma_t t}) \quad , \qquad (3.10b)$$

where $v_0$ is the terminal velocity:

$$v_0 = \frac{g(\sin\theta - \mu\cos\theta)}{\gamma_t} \quad . \qquad (3.10c)$$

Tangential coefficient of damping $\gamma_t$ is calculated using Eq. (3.10b) with the observed velocity, or Eq. (3.10c) with the terminal velocity. The blocks' terminal sliding velocity was found to be $v_0$=(0.43±0.08)m/s. Assuming equality of the dynamic and static coefficients of friction (and given by $\mu$=0.43±0.3), we can use Eq. (3.10c) to obtain $\gamma_t$=(3.0±0.5)s$^{-1}$.

### 3.5 Density of wood particles (ρ)

Various methods for measuring particle density have been reported [17]. Since our particles were prismatic, a particle's volume was determined directly from its dimensions, its mass by using a precision balance, and its density as simply mass/volume. Sampling five particles, a mean density 0.45×10$^{-3}$ g/mm$^3$ was found (standard deviation 6%). Since the length of the blocks was 120mm, the surface density of the particles was 54×10$^{-3}$ gm/mm$^2$.

| Parameter | Units | Parameter Name | Value | Comment |
|---|---|---|---|---|
| $k_n$ | N/mm | Normal Stiffness | $1 \times 10^4$ | Experimentally calculated using force-displacement relation |
| $k_t$ | N/mm | Tangential Stiffness | $6.5 \times 10^2$ | Numerically calculated using elastic analysis of two particles in contact |
| $\mu_{wall}$ | Dimensionless | Particle-wall Coefficient of friction | 0.44 | Experimentally calculated using a sliding test. |
| $\mu_{particle}$ | Dimensionless | Particle-particle Coefficient of friction | 0.38 | Experimentally calculated using a sliding test. |
| $\gamma_n^{p-w}$ | 1/s | Particle-wall Coefficient of Damping | 4303 | Experimentally calculated using pendulum test for coefficient of restitution |
| $\gamma_n^{p-p}$ | 1/s | Particle-particle Coefficient of Damping | 6085 | Experimentally calculated using dropping test for coefficient of restitution |
| $\gamma_t$ | 1/s | Tangential Coefficient of Damping | 3.0 | Estimated from sliding test via terminal velocity |

**Table 3.4 Summary of the determination of the DEM parameters.**

# 4 Experimental approach

In the experiment we used a model silo: 500mm in height and 300mm in width made of 3mm-thick methacrylate plates. The geometry of the hopper was modifiable. We used sixteen different configurations, with angles of 0°, 30°, 45°, and 60°, and outlet openings of 40mm, 60mm, 100mm, and 150mm (Figure 4.1).

In each test the silo was filled with prismatic particles of nineteen different cross-sectional shapes (Figure 4.2). The primitive shapes are numbered 12, 13, 14, and 15; the rest were obtained from these by bevelling the corners. A high-speed camera was set in front of the silo and recorded the whole process of discharge, from the time the bottom gate was opened until the silo was empty or clogged. It was ensured that the particles inside the silo never touched the wall behind. For diameter 100mm, the trials were repeated three or four times and discharge times were averaged; and similarly for diameter 150mm.

To observe flow patterns, half of the wood particles were painted on their front surface. The particles were placed in alternating painted and non-painted layers of width 45mm.

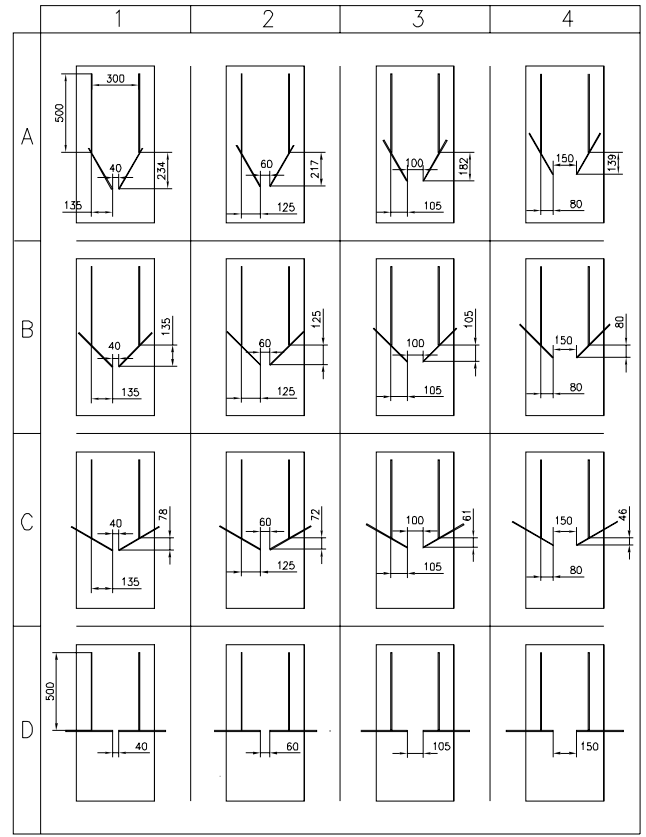

**Figure 4.1 Geometrical configurations of the model silo (dimensions in mm).**

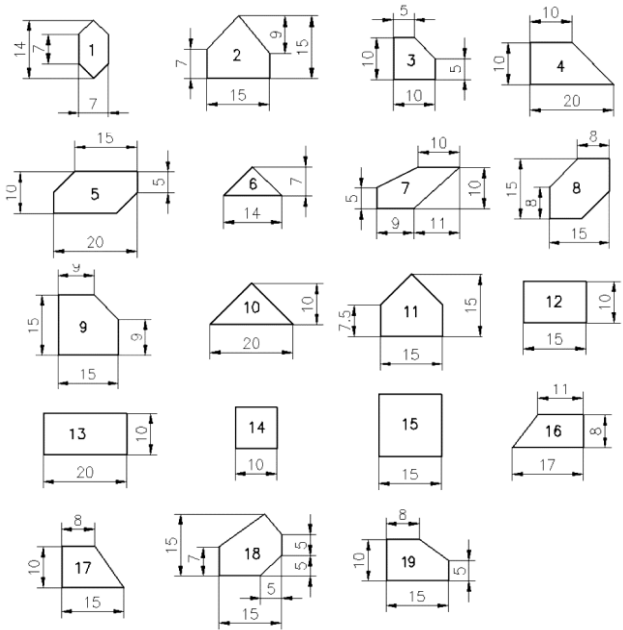

**Figure 4.2 Cross-sectional shapes of wood particles (dimensions in mm**

## 4.1 Experimental discharge time

Discharge was begun by manually removing the bottom cover, so that particles flowed under gravity. With diameters of 40mm and 60mm the aperture was found to be insufficient, clogging after only a few particles had fallen (Figure 4.3). Arches consisting of up to 16 particles were then found, contrasting with 2D-flow spherical particle experiments where arches up to 12 particle diameters are reported [15].

Whenever clogging was observed, trials were repeated to determine its frequency. At hopper angles of 60° and 45°, the frequency of clogging was 20% and 25% respectively. With 100mm outlet diameter there were brief interruptions to particle flow.

With outlet diameter 150mm, the particles flowed freely. These trials were run three times, and average discharge times are shown in Table 4.1. Standard deviations were below 5%, small enough to consider the experimental results accurate. Silo D, with the outlet set at angle 0° (horizontal at the bottom), did not empty completely, as shown in Figure 4.3. However, silos A, B, and C – set at angles 60°, 45°, and 30° respectively – did empty. The time for emptying was recorded. The trends were that discharge times shortened as aperture increased, and also as angle from horizontal increased.

| Silo | Θ | Exp | Aperture | | | |
|---|---|---|---|---|---|---|
| | | | 1 | 2 | 3 | 4 |
| | | | 40 mm | 60 Mm | 100 mm | 150 mm |
| A | 60 | 1 | C | C | 2.15 | 0.96 |
| | | 2 | | | C | 1 |
| | | 3 | | | 2.1 | 1.01 |
| | | 4 | | | 2.05 | |
| | | 5 | | | 207 | |
| | Mean | | | | **2.09** | **0.99** |
| | Std | | | | 1.9% | 3.0% |
| B | 45 | 1 | C | C | 2.42 | 1.28 |
| | | 2 | | | C | 1.19 |
| | | 3 | | | 2.79 | 1.16 |
| | | 4 | | | 2.55 | |
| | Mean | | | | **2.59** | **1.21** |
| | Std | | | | 7.3% | 5.0% |
| C | 30 | 1 | C | C | 2.8 | 1.51 |
| | | 2 | | | 3.11 | 1.53 |
| | | 3 | | | 3.28 | 1.42 |
| | Mean | | | | **3.06** | **1.49** |
| | Std | | | | 7.8% | 4.0% |
| D | 0 | 1 | C | C | 2.78 | 2.2 |
| | | 2 | | | 2.38 | 1.79 |
| | | 3 | | | 2.68 | 1.77 |
| | Mean | | | | **2.61** | **1.92** |
| | Std | | | | 8.0% | 12.6% |

**Table 4.2 Recorded discharge times (in seconds) in the experiments. The letter C indicates clogging.**

## 4.2 Comparison of experiment with DEM results

The experiments with all silos shown in Figure 4.1 were modelled using spheropoly-hedra of prismatic shape, and cross sections as shown in Figure 4.2. A spheroradius of 1mm was used to represent the roundness of the particles. The particles were allowed to rotate only about their longitudinal axis. This was a reasonable approximation – except for the last discharged particles, which underwent 3D rotation as they slid on the hopper.

The first step in the DEM simulations was to emulate hand-filling of the silos. To achieve similar configurations, particles were poured in at the top of the silo and the friction was set to zero – avoiding creation of large voids between particles, which were not observed in the experiment. As the second step, the silo was opened by removing the bottom door, allowing the particles to be discharged under gravity.

Final configurations after discharge for DEM simulations are shown in Figure 4.4. They are similar to the experimental results in Figure 4.3. In both, clogging occurred with outlet diameters set to 40mm and 60mm. With 100mm diameter, interrupted and intermittent flow was observed in both the experimental regime and the simulation regime. With 150mm outlet diameter, A, B, and C silos all emptied in both regimes. For the rectangular silo with 150mm outlet diameter (called D4), some particles remained on the sides of the exit both in experiments and in simulations.

We also found good agreement in discharge times. A summary of the experimentally recorded times from Table 4.1 is presented in Table 4.2. Times calculated from the DEM simulations are also summarised in this table; they are close to the experimental results. For the silo labelled as C3 with 100mm outlet diameter, experimental and simulated discharge times were 3.06s and 3.05s respectively. The discrepancy was attributed to the relatively short time of discharge (around a few seconds) and difficulties in capturing the moment when the last particle left the exit, due to its off-plane rotation during the experiments

| Angle | Letter | Experimental | Numerical |
|---|---|---|---|
| **60** | **A** | 0.99 | 1.00 |
| **45** | **B** | 1.21 | 1.19 |
| **30** | **C** | 1.49 | 1.64 |
| **0** | **D** | 1.92 | 1.78 |

**Table 4.2 Comparison of experimental and numerical discharge times (in seconds).**

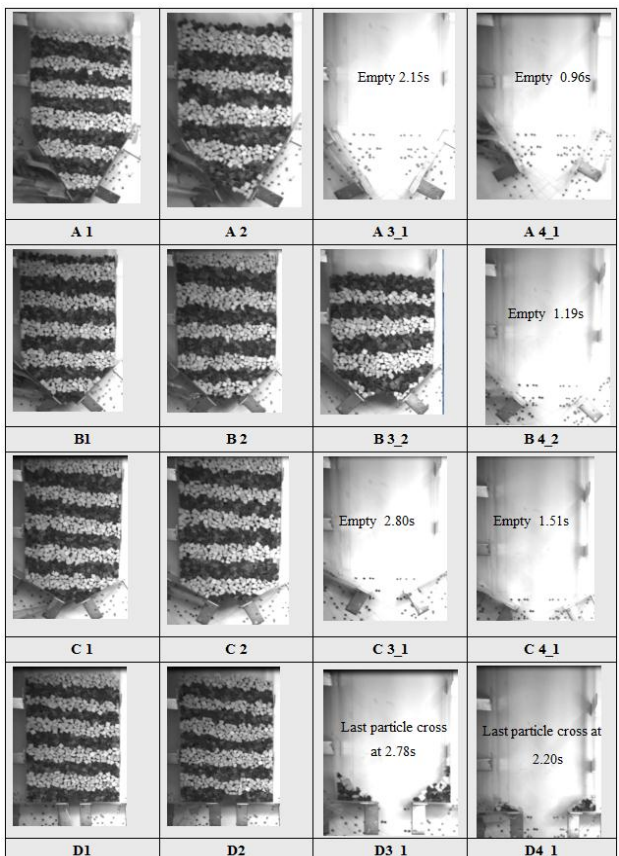

**Figure 4.3 Final silo configurations in experiments.**

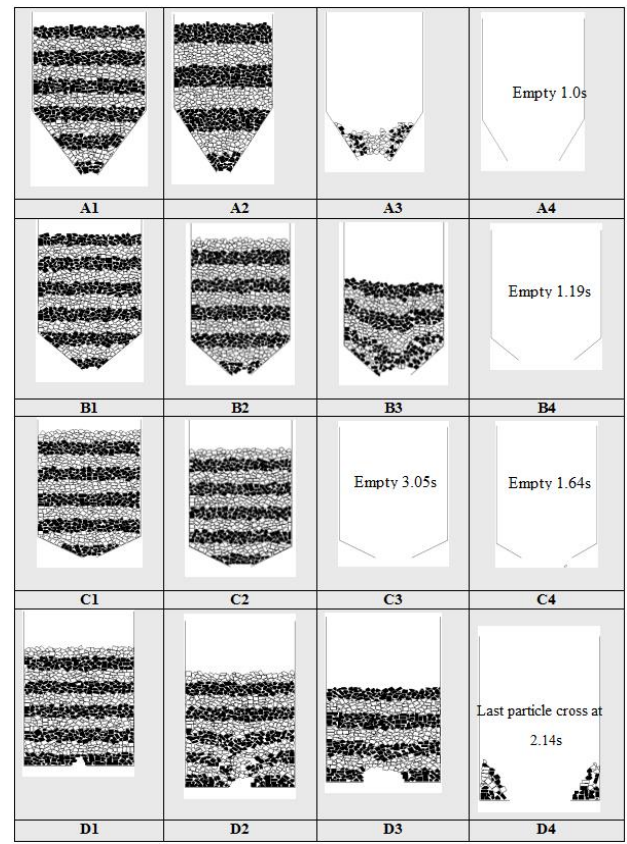

**Figure 4.4 Final configurations in DEM simulations.**

### 4.2.1 Displacement profiles

Particles' displacement profiles are very similar in the experiments and the DEM simulations. A typical comparison is shown in Figures 4.5 and 4.6 for silo A4. The displacement profiles are shown at 0.25 seconds intervals. A V-shape in the displacement profile is observed in both regimes, with a certain mix of the layers near the exit due to eddy-like deformation of groups of particles as they flow. Intermittent flow and slip-stick motion at the walls were observed both in experiments and in simulations.

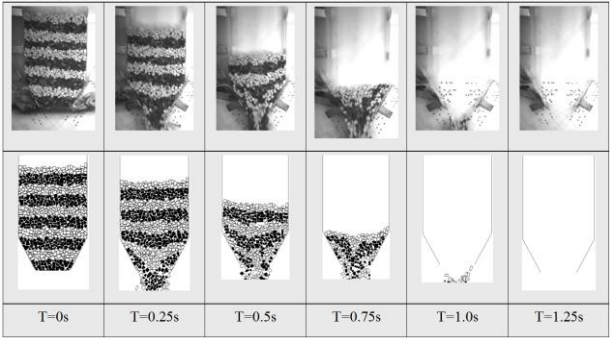

**Figure 4.5 Snapshots in the experiments (top) and simulation (bottom) in the A4 silo.**

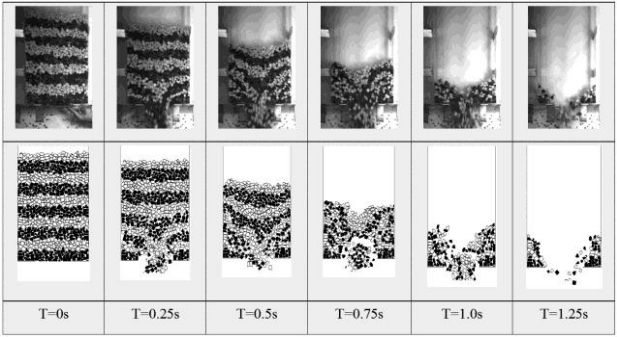

**Figure 4.6 Snapshots in the experiments (top) and simulation (bottom) in the D4 silo.**

### 4.3 DEM sensitivity

#### 4.3.1 $k_n$ sensitivity

To analyse the sensitivity of the simulations with respect to $k_n$, a series of simulations were performed with $k_n$ varied but $k_t/k_n$ and $\gamma_n/\omega_0$ held constant. The results, shown in Table 4.4, might suggest that decreasing the value of $k_n$ will slightly increase the discharge time. This result may be useful to speed the simulation: decreasing $k_n$ makes the particles less stiff, so that time step in the simulations can be increased without affecting the simulation time much. With moderately small values of $k_n$ the discharge times are not strongly affected; but with times such as $1 \times 10^3$ N/mm, overlapping between the spheropolygons is unrealistic, and contact forces are unreliable.

| $k_n$ (N/mm) | $\gamma_n$ | $\varepsilon$ | $t$ (sec) |
|---|---|---|---|
| $6.25 \times 10^3$ | $1.00 \times 10^3$ | 0.54 | 0.99 |
| $2.50 \times 10^3$ | $2.00 \times 10^3$ | 0.54 | 0.99 |
| $1.00 \times 10^4$ | $4.00 \times 10^3$ | 0.54 | 0.99 |
| $4.00 \times 10^4$ | $8.00 \times 10^3$ | 0.54 | 0.98 |

**Table 4.4 Sensitivity of $k_n$.**

Similarly the variation of time with $k_t$, shown in Table 4.5, is insignificant as well. It is also clear from the results that if the $k_t$ value increases the discharge time increases.

| $k_t$ (N/mm) | $k_t / k_n$ | $t$ (sec) |
|---|---|---|
| $0.50 \times 10^3$ | 0.05 | 0.96 |
| $0.75 \times 10^3$ | 0.075 | 1.00 |
| $1.00 \times 10^3$ | 0.10 | 0.99 |
| $2.50 \times 10^3$ | 0.25 | 0.99 |
| $3.75 \times 10^3$ | 0.375 | 1.02 |
| $5.00 \times 10^3$ | 0.50 | 1.04 |

**Table 4.5 Sensitivity of $k_t$.**

Sensitivity analysis was also conducted with different values of the coefficient of friction (**μ**). As shown in Table 4.6, small changes in **μ** affected the time significantly more than variation in other parameters.

| $\mu$ | $t$ (sec) |
|---|---|
| 0.10 | 0.70 |
| 0.20 | 0.80 |
| 0.30 | 0.86 |
| 0.40 | 0.99 |
| 0.50 | 1.08 |

**Table 4.6 Sensitivity of coefficient of friction.**

The restitution coefficient was changed by varying the normal coefficient of damping. As damping is increased, discharge time increases

(see Table 4.7). The sensitivity of the flow rate to changes in the normal coefficient of damping $\gamma_n$ is low, with the time varying by less than 8% for a change in $\gamma_n$ from 1000 to 16000 s$^{-1}$.

| $\gamma_n$ (s$^{-1}$) | $\varepsilon$ | $t$ (sec) |
|---|---|---|
| 1000 | 0.86 | 0.93 |
| 2000 | 0.74 | 0.99 |
| 4000 | 0.54 | 0.99 |
| 8000 | 0.27 | 1.02 |
| 16000 | 0.022 | 1.05 |

**Table 4.7 Sensitivity of normal coefficient of damping $\gamma_n$.**

The sensitivity of the flow rate to changes in the tangential coefficient of damping $\gamma_t$ is quite low. The coefficient could have been taken as zero without significant effects on the flow rate.

| $\gamma_t$ (s$^{-1}$) | $t$ (sec) |
|---|---|
| 0 | 0.99 |
| 3 | 0.99 |
| 10 | 1.02 |
| 100 | 1.13 |

**Table 4.8 Sensitivity of tangential coefficient of damping $\gamma_t$.**

## 5 Conclusions

The discrete spheropolyhedron-based models have produced improved simulations in terms of particle shape and multiple-contact interactions. The next steps brought forward in this work were establishment of methods for determination of relevant parameters, and validation of the model experimentally with particles in silos. Flow rates and displacement patterns from our numerical simulations were very similar those in the experiments. Differences between 2.1% and 4.7% in discharge times are small, taking into account the shortness of those discharge times.

Clearly, flow rate of particles increased with the outlet diameter. In both simulations and experiments it is observed also to increase with the angle of the hopper to the horizontal. Comparisons of discharge times between experiments and simulations establish high levels of accuracy.

Several challenges remain, in determining discrete element modelling parameters. One difficulty was the non-linearity and variability of normal stiffness ($k_n$), partially from the effective stiffness at contact which is lower than bulk material stiffness, and from randomness in surface roughness features at particle contacts. Nonetheless, discharge time was largely insensitive to contact stiffness, and we were able to reproduce experimental values using a constant value. However, we presume these non-linearities may need to be taken into account for other scenarios, such as loading of confined granular materials. In these cases determination of the contact stiffness resulting from surface-asperity flattening would allow accurate simulation of particle-particle interactions where deformation occurs only in surface structures.

Another challenge was the experimental calculation of tangential stiffness due to the existence of contact moments, whose values were not accessible in the laboratory. To obtain a measure of tangential stiffness, we used finite element simulations of two blocks in contact. We found that $k_t/k_n$ varied from 0.1 to 0.25, depending on whether we introduce the topography of the contact surface in the finite element method. However, changing the tangential stiffness without this ratio had only a minor influence in the discharge flow.

We found the main parameters controlling flow rate to be coefficient of friction and to a lesser extent the restitution coefficients. Damping forces during sliding were small, and they could have been ignored for the simulations.

Our analysis of contact stiffness shows the complexity of the load-deformation response, where non-linearity and fluctuations due to surface roughness are salient. Yet it is quite remarkable from our simulations that stiffness plays no essential role in determining the mass flow. This is consistent with the general trend in physics research: less focus on stiffness and restitution, and more on the nature of frictional forces. One might even argue that attention to parameters of contact stiffness is not required in an accurate predictive model for granular flow. We recommend obtaining friction and restitution coefficients from inclined plane tests, and pendulum and dropping tests, repeating the experiment on several particles but avoiding the expense of determining other DEM parameters.

Validation studies for granular flow might also be enhanced by going beyond single (usually mean) values in simulating mechanical properties. Contacts with values drawn from experimental distributions, perhaps. More realistic flow has been already observed when particle size distributions are used, instead of monodisperse systems where the systems tend to crystallize. An open question to explore in future simulations: What is the effect of disorder in the contact material parameters on the mass flow properties?

Although this research studied discharge times and used visual observations, other methods were available for quantitative support of the results obtained – such as determining velocity profiles or residential times of particles at different levels in the silo, or determination of the mass flow index. Any of these methods would have lent force to our conclusions.

# 6 Acknowledgments

The authors acknowledge technical support from Ross Barker and Shiao-Huey Chow in the PIV analysis of interface deformation tests. We sincerely thank David Airey, Jørgen Nielsen, and Celia Lozano Grijalba for their helpful discussions with us. FAM is supported by the CERDS funding scheme.

# Appendix (for review purposes)

*Relating coefficient of restitution with $\gamma_n$*

The coefficients of restitution are related with the normal coefficient of damping $\gamma_n$ using an analytical derivation of the collisions between two blocks. Let us assume that a block of mass $m_1$ impacts a second one of mass $m_2$. The positions of the blocks satisfy the differential equations

$$m_1 \frac{d^2 x_1}{dt^2} = -k_n(x_1 - x_2) - m_{12}\gamma_n(v_1 - v_2) \quad (3.7)$$

$$m_2 \frac{d^2 x_2}{dt^2} = k_n(x_1 - x_2) + m_{12}\gamma_n(v_1 - v_2) \quad (3.8)$$

where effective mass $m_{12} = m_1 m_2/(m_1 + m_2)$ (3.9)

Subtracting Eq. (3.7) from Eq. (3.8), we derive

$$\frac{d^2(x_1 - x_2)}{dt^2} = -k_n \delta \left(\frac{1}{m_1} + \frac{1}{m_2}\right) - m_{12}\gamma_n(v_1 - v_2)\left(\frac{1}{m_1} + \frac{1}{m_2}\right) \quad (3.10)$$

Rearrange Eq.(3.10) we get

$$\frac{d^2 \delta}{dt^2} + \gamma_n \frac{d\delta}{dt} + \omega_0^2 \delta = 0 \qquad \omega_0 = \sqrt{\frac{k_n}{m_{12}}} \qquad \delta = x_1 - x_2 \quad (3.11)$$

This equation is solved with initial conditions $\delta = 0$ and $d\delta/dt = v_0$ at $t = 0$, where $v_0$ is the relative velocity before impact.

$$v_f = -v_0 \exp\left(-\frac{\gamma_n \pi}{2\omega}\right) \qquad \omega = \sqrt{\omega_0^2 - \left(\frac{\gamma_n}{2}\right)^2} \quad (3.12)$$

If $v_f$ is the relative velocity after impact, the analytical solution of the restitution coefficient is

$$\varepsilon = \frac{-v_f}{v_0} = \frac{v_0 \exp\left(-\frac{\gamma_n}{2}\frac{\pi}{\omega}\right)}{v_0} = \exp\left(-\frac{\gamma_n}{2}\frac{\pi}{\omega}\right) \quad (3.13)$$

To find the final velocities of the two particles $v_1$ and $v_2$, Eqs. (3.14) and (3.15) are simultaneously solved. Equation 3.14 means that the difference between the two blocks' final velocities is the relative velocity after impact. Eq. 3.15 is derived by considering momentum conservation of the system.

$$v_1 - v_2 = v_f = -v_0 \varepsilon \quad (3.14)$$

$$m_1 v_0 = m_1 v_1 + m_2 v_2 \quad (3.15)$$

$$v_1 = v_0 \frac{\varepsilon m_1 - m_2}{m_1 + m_2} \quad (3.16)$$

$$v_2 = v_0 (1+\varepsilon) \frac{m_1}{m_1 + m_2} \quad (3.17)$$

In the case of particle-particle interaction, $m_1 = m_2$ and $m_{12} = m_1$. Eqs. 3.13, 3.16 and 3.17 become Eqs. 3.18, 3.19 and 3.20, respectively.

$$\varepsilon_p = \exp\left(-\frac{\gamma_n}{2} \frac{\pi}{\sqrt{\frac{2k_n}{m_1} - \left(\frac{\gamma_n}{2}\right)^2}}\right) \quad (3.18)$$

$$v_1 = v_0 \frac{\varepsilon - 1}{2} \tag{3.19}$$

$$v_2 = \frac{v_0(1+\varepsilon)}{2} \tag{3.20}$$

For the case of the particle-wall interaction the mass of the wall is much larger than the mass of the particle so that $m_2$ is infinite and $m_{12} = m_1$. The Eqs.3.13, 3.16 and 3.17 become 3.21, 3.22 and 3.23 respectively.

$$\varepsilon_w = \exp\left(-\frac{\gamma_n}{2} \frac{\pi}{\sqrt{\frac{k_n}{m_1} - \left(\frac{\gamma_n}{2}\right)^2}}\right) \tag{3.21}$$

$$v_1 = -v_0 \varepsilon \tag{3.22}$$

$$v_2 = 0 \tag{3.23}$$